\documentclass[twocolumn,amssymb, amsmath, aps, superscriptaddress, showpacs, footinbib,  prb]{revtex4}
\usepackage{graphicx}

\newcommand{\eff}{\text{eff}}
\newcommand{\AFM}{\text{AFM}}
\newcommand{\FM}{\text{FM}}
\newcommand{\CW}{\text{CW}}
\newcommand{\ph}{\text{phon}}
\newcommand{\mg}{\text{mag}}
\newcommand{\reg}{\text{reg}}
\newcommand{\eps}{\varepsilon}

\begin{document}

\title{Frustrated square lattice with spatial anisotropy:\\ crystal structure and magnetic properties of PbZnVO(PO$_4)_2$}
\author{Alexander A. Tsirlin}
\email{altsirlin@gmail.com}
\affiliation{Max Planck Institute for Chemical Physics of Solids, N\"{o}thnitzer
Str. 40, 01187 Dresden, Germany}
\affiliation{Department of Chemistry, Moscow State University, 119991 Moscow, Russia}
\author{Ramesh Nath}
\altaffiliation[Present address: ]{Indian Institute of Science Education and Research,
Trivandrum-695016 Kerala, India}
\affiliation{Max Planck Institute for Chemical Physics of Solids, N\"{o}thnitzer
Str. 40, 01187 Dresden, Germany}
\author{Artem M. Abakumov}
\affiliation{Department of Chemistry, Moscow State University, 119991 Moscow, Russia}
\affiliation{EMAT, University of Antwerp, Groenenborgerlaan 171, B-2020 Antwerp, Belgium}
\author{Roman V. Shpanchenko}
\affiliation{Department of Chemistry, Moscow State University, 119991 Moscow, Russia}
\author{Christoph Geibel}
\author{Helge Rosner}
\email{Helge.Rosner@cpfs.mpg.de}
\affiliation{Max Planck Institute for Chemical Physics of Solids, N\"{o}thnitzer
Str. 40, 01187 Dresden, Germany}

\begin{abstract}
Crystal structure and magnetic properties of the layered vanadium phosphate PbZnVO(PO$_4)_2$ are studied using x-ray powder diffraction, magnetization and specific heat measurements, as well as band structure calculations. The compound resembles AA$'$VO(PO$_4)_2$ vanadium phosphates and fits to the extended frustrated square lattice model with the couplings $J_1,J_1'$ between nearest-neighbors and $J_2,J_2'$ between next-nearest-neighbors. The temperature dependence of the magnetization yields estimates of averaged nearest-neighbor and next-nearest-neighbor couplings, $\bar J_1\simeq -5.2$~K and $\bar J_2\simeq 10.0$~K, respectively. The effective frustration ratio $\alpha=\bar J_2/\bar J_1$ amounts to $-1.9$ and suggests columnar antiferromagnetic ordering in PbZnVO(PO$_4)_2$. Specific heat data support the estimates of $\bar J_1$ and $\bar J_2$ and indicate a likely magnetic ordering transition at 3.9~K. However, the averaged couplings underestimate the saturation field, thus pointing to the spatial anisotropy of the nearest-neighbor interactions. Band structure calculations confirm the identification of ferromagnetic $J_1,J_1'$ and antiferromagnetic $J_2,J_2'$ in PbZnVO(PO$_4)_2$ and yield $(J_1'-J_1)\simeq 1.1$~K in excellent agreement with the experimental value of $1.1$~K, deduced from the difference between the expected and experimentally measured saturation fields. Based on the comparison of layered vanadium phosphates with different metal cations, we show that a moderate spatial anisotropy of the frustrated square lattice has minor influence on the thermodynamic properties of the model. We discuss relevant geometrical parameters, controlling the exchange interactions in these compounds, and propose a new route towards strongly frustrated square lattice materials.
\end{abstract}

\maketitle

\section{Introduction}
Quantum magnetism is one of the active research topics in solid state physics. Quantum effects lead to numerous unusual properties, interesting with respect to the fundamental understanding of critical phenomena (spin-liquid ground states, Bose-Einstein condensation in high magnetic fields)\cite{giamarchi,lee} and potential technological applications (magnetoelectric coupling, ballistic heat transport).\cite{licu2o2,heat-transport} The search for new quantum magnets that enable to challenge theoretical predictions remains a long-standing problem in solid state science, because theoretical studies usually deal with relatively simple models, like the Heisenberg model, while the properties of real compounds are often determined by the interplay of numerous effects, such as isotropic and anisotropic exchange interactions, orbital and/or lattice degrees of freedom, etc. 

The frustrated square lattice (FSL) with isotropic exchange is an archetypal spin model in two dimensions.\cite{misguich} This model reveals strong quantum fluctuations due to the combination of low-dimensionality and frustration. The specific regime of the model is determined by the frustration ratio, i.e., by the ratio of the next-nearest-neighbor (NNN) interaction ($J_2$) to the nearest-neighbor (NN) interaction ($J_1$): $\alpha=J_2/J_1$. The frustration ratio determines the ground state of the system -- either ordered [N\'eel antiferromagnetic, columnar antiferromagnetic, or ferromagnetic (FM) ordering] or disordered.\cite{shannon2004,darradi2008} The precise nature of the disordered [presumably spin-liquid at $J_2/J_1\simeq 0.5$ (Ref.~\onlinecite{misguich}) and nematic at $J_2/J_1\simeq -0.5$ (Ref.~\onlinecite{shannon2006})] ground state remains controversial, because a theoretical treatment of the model is inevitably approximate.

Experimental studies aim at the search for materials that fit to the FSL model. Recently, extensive investigations identified a number of vanadium-based FSL compounds. Li$_2$VOSiO$_4$ and Li$_2$VOGeO$_4$ reveal antiferromagnetic (AFM) $J_1$ and $J_2$ with $J_2\gg J_1$.\cite{melzi2000,rosner2002,HTSE} Thus, the columnar AFM ordering is established, while the frustration is weak.\cite{bombardi2004} In the $J_1>J_2$ (both $J_1$ and $J_2$ AFM) region, VOMoO$_4$ undergoes N\'eel-type ordering,\cite{carretta2002,bombardi2005} yet the presence of the long-range order in PbVO$_3$ is still under debate.\cite{tsirlin2008,oka2008} The region of FM $J_1$ -- AFM $J_2$ is probed by the family of layered vanadium phosphates AA$'$VO(PO$_4)_2$ (AA$'$ = Pb$_2$, BaZn, SrZn, and BaCd).\cite{kaul2004,enrique,nath2008,carretta2009} None of the latter compounds fall in the critical region with $J_2/J_1\simeq -0.5$, although two of them, BaCdVO(PO$_4)_2$ and SrZnVO(PO$_4)_2$, reveal $\alpha\simeq -1$ and show a pronounced effect of the frustration on the thermodynamic properties.\cite{enrique,nath2008} An additional feature of the layered vanadium phosphates is the lack of the tetragonal symmetry and the resulting spatial anisotropy of the spin lattice (both for the NN and NNN couplings).\cite{tsirlin2009}

Motivated by the above-mentioned demand for strongly frustrated FSL materials, we attempted to extend the family of layered vanadium phosphates. Metal cations accommodated between the V--P--O layers do not take part in the magnetic exchange. However, the size of these cations determines the geometry of the magnetic layer,\cite{tsirlin2009} hence the replacement of metal cations can be a promising way to tune the spin system of the material. The approach of cation substitution is widely used in solid state chemistry. For example, the replacement of Li by Na in LiVO$_2$ leads to a change in the orbital ordering pattern and results in the long-range magnetic ordering in NaVO$_2$ (Ref.~\onlinecite{navo2}) instead of the formation of trimer clusters in LiVO$_2$.\cite{livo2} Following a similar approach for the FSL compounds, we succeeded in preparing the vanadium phosphate PbZnVO(PO$_4)_2$ that belongs to the family of layered AA$'$VO(PO$_4)_2$ vanadium phosphates and combines certain structural features of SrZnVO(PO$_4)_2$ and BaZnVO(PO$_4)_2$. To the best of our knowledge, neither the preparation and the crystal structure, nor the magnetic properties of PbZnVO(PO$_4)_2$ have been reported before. The only exception is the high-field magnetization curve of PbZnVO(PO$_4)_2$ presented in the comparative study of high-field properties of the FSL compounds.\cite{high-field}

In the following, we present the results of a combined -- structural, phenomenological, and microscopic -- study of the model FSL compound PbZnVO(PO$_4)_2$. We start with the methodological aspects in Sec.~\ref{methods}. In Sec.~\ref{structure}, the structural data are reported. Sec.~\ref{experiment} deals with the experimental study of the magnetic properties, while Sec.~\ref{band} presents band structure calculations and the evaluation of individual exchange couplings. In Sec.~\ref{modeling}, we further exploit the computational approach and consider the influence of different structural factors on the magnetic interactions in layered vanadium phosphates. Sec.~\ref{discussion} concludes the study with the discussion and a summary.

\section{Methods}
\label{methods}
Polycrystalline samples of PbZnVO(PO$_4)_2$ were obtained by heating a mixture of PbZnP$_2$O$_7$, V$_2$O$_3$, and V$_2$O$_5$ in an evacuated quartz tube ($10^{-2}$ mbar) at 700~$^{\circ}$C for 24 hours. Phase composition of the samples were controlled using x-ray diffraction (Huber G670 Guinier camera, CuK$_{\alpha1}$ radiation, $2\theta=3-100^{\circ}$ angle range). PbZnP$_2$O$_7$ was obtained by heating a stoichiometric mixture of PbO, ZnO, and NH$_4$H$_2$PO$_4$ in air at 750~$^{\circ}$C for 48 hours. 

The best sample was obtained from the reactant mixture with the stoichiometric cation composition and the slight oxygen excess corresponding to the PbZnVO$_{1.09}$(PO$_4)_2$ formula. The sample contained the targeted PbZnVO(PO$_4)_2$ phase and the minor impurity of diamagnetic PbZnP$_2$O$_7$ (about 2 wt.~\%). This sample was further used for the structure refinement and for the thermodynamic measurements. According to the structure refinement and magnetization measurements, the PbZnVO(PO$_4)_2$ compound is stoichiometric.\cite{note2} The annealing of the PbZnVO(PO$_4)_2$ samples above 700~$^{\circ}$C resulted in the decomposition of the compound.

The x-ray powder pattern for the structure refinement was collected using the STOE STADI-P diffractometer (transmission geometry, Ge(111) monochromator, CuK$_{\alpha1}$ radiation, linear position-sensitive detector, angle range $2\theta=7-100^{\circ}$). The structure refinement was performed using the JANA2000 program.\cite{jana2000}

Temperature dependence of the magnetization was measured using the Quantum Design MPMS SQUID in the temperature range of $2-360$~K. Heat capacity data were collected with the commercial PPMS setup. To improve the contacts between the grains, the pellet for the heat capacity measurement was additionally annealed overnight in the evacuated quartz tube at 500~$^{\circ}$C. The x-ray study of the re-annealed sample evidenced that the phase composition remained unchanged. 

Scalar-relativistic band structure calculations were performed using the full-potential local-orbital scheme (FPLO8.50-32) within the local density approximation (LDA) of density functional theory.\cite{fplo} The Perdew-Wang version of the exchange-correlation potential was applied.\cite{perdew-wang} The calculations were performed for the crystallographic unit cell of PbZnVO(PO$_4)_2$ (112 atoms, $k$ mesh of 256 points with 75 points in the irreducible part) and for a number of model structures with the formal composition of LiVOPO$_4$ (32 atoms, $k$ mesh of 4096 points with 729 points in the irreducible part). The construction of the model structures is further described in Ref.~\onlinecite{tsirlin2009}. LDA band structures were used to select the relevant states and to analyze the respective bands with tight-binding (TB) models. The parameters of these models were evaluated as overlap integrals of Wannier functions localized on vanadium sites\cite{wannier,wannier2} and further used to estimate exchange integrals. Details of the procedure are given in Sec.~\ref{band}.

\section{Crystal structure}
\label{structure}
The X-ray powder pattern of PbZnVO(PO$_4)_2$ was indexed in an orthorhombic unit cell with lattice parameters $a=8.763(1)$~\r A, $b=9.072(1)$~\r A, \mbox{$c=18.070(5)$~\r A}. This unit cell is similar to unit cells of the compositionally-related compounds BaZnVO(PO$_4)_2$ ($a=8.814$~\r A, $b=9.039$~\r A, $c=18.538$~\r A)\cite{baznvp2o9} and SrZnVO(PO$_4)_2$ ($a=9.066$~\r A, $b=9.012$~\r A, $c=17.513$~\r A).\cite{srznvp2o9} Reflection conditions $(hk0)$, \mbox{$h=2n$}; $(h0l), l=2n; (0kl), k=2n$ unambiguously pointed to the $Pbca$ space group and supported the similarity to the AA$'$VO(PO$_4)_2$ layered vanadium phosphates (AA$'$ = BaZn, SrZn, BaCd). To select the proper starting model for the structure refinement, we simulated powder patterns using lattice parameters of PbZnVO(PO$_4)_2$, atomic positions of BaZnVO(PO$_4)_2$ or SrZnVO(PO$_4)_2$, and Pb atom instead of Ba/Sr. The structural data for BaZnVO(PO$_4)_2$ showed better agreement with the experimental pattern. Therefore, the structure of BaZnVO(PO$_4)_2$ was used as a starting model for the refinement.

\begin{figure}
\includegraphics{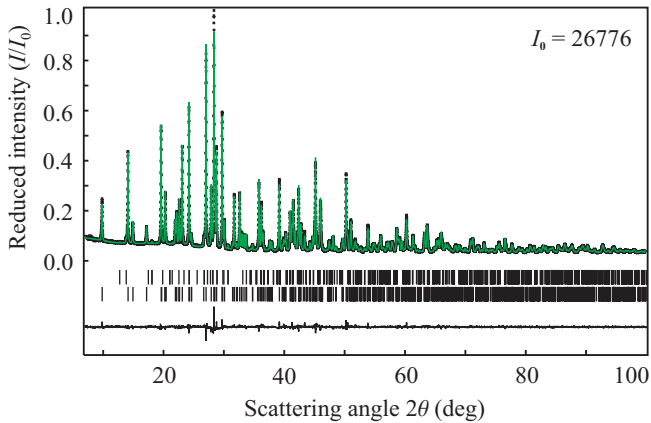}
\caption{\label{refinement}
(Color online)
Structure refinement for PbZnVO(PO$_4)_2$: experimental, calculated, and difference X-ray diffraction patterns. Upper and lower sets of ticks show line positions for the PbZnP$_2$O$_7$ impurity and for PbZnVO(PO$_4)_2$, respectively.
}
\end{figure}

For the final refinement, the PbZnP$_2$O$_7$ impurity was taken into account.\cite{pbznp2o7} Atomic displacement parameters for two positions of phosphorous and for nine positions of oxygen were independently constrained due to the huge difference in the scattering powers of Pb and the light O and P atoms. The refinement converged to $R_I=0.019,R_p=0.028$, and $\chi^2=2.55$. The experimental, simulated, and difference diffraction patterns are shown in Fig.~\ref{refinement}. Atomic positions are listed in Table~\ref{coordinates}, while interatomic distances, relevant for a further discussion of the magnetic interactions, are presented in Table~\ref{distances}.

\begin{figure*}
\includegraphics{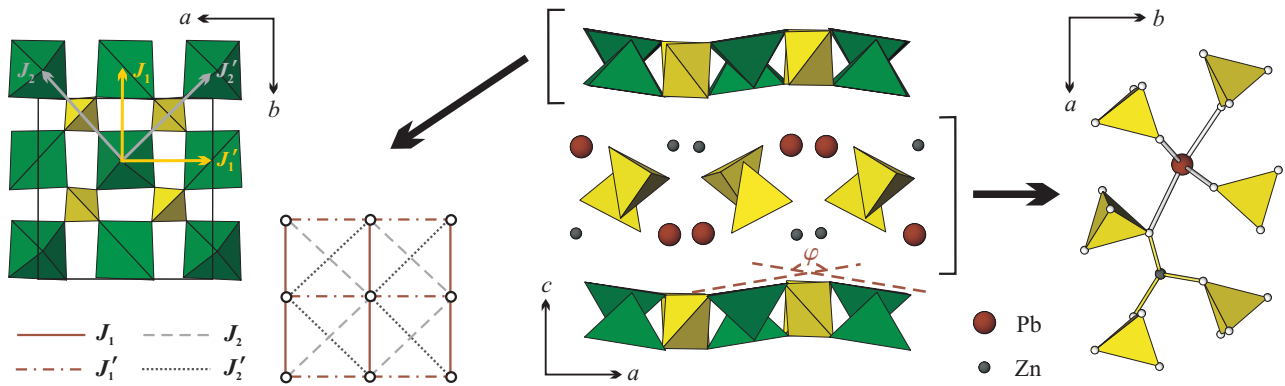}
\caption{\label{fig_structure}
(Color online) Crystal structure and spin model of PbZnVO(PO$_4)_2$: the [VOPO$_4$] layer and the magnetic interactions (left panel); overall view of the crystal structure (middle panel); bonding of Pb and Zn cations in the interlayer blocks (right panel). Larger and smaller spheres denote Pb and Zn. The $\varphi$ angle measures the buckling of the [VOPO$_4$] layers. In the sketch of the spin lattice (left panel), solid, dash-dotted, dashed, and dotted lines show the interactions $J_1$, $J_1'$, $J_2$, and $J_2'$, respectively.
}
\end{figure*}

The crystal structure of PbZnVO(PO$_4)_2$ is shown in Fig.~\ref{fig_structure}. Vanadium atoms form distorted octahedra with the short vanadyl bond of 1.55(1)~\r A, four longer bonds of $2.00-2.13$~\r A in the equatorial plane, and a rather long bond of 2.38(1)~\r A opposite to the vanadyl one. If the latter atom is cast away, the vanadium coordination is reduced to a square pyramid (in the following, we will refer to the square-pyramidal coordination of vanadium in order to simplify the comparison to the other AA$'$VO(PO$_4)_2$ compounds). Phosphorous atoms reside in slightly distorted PO$_4$ tetrahedra typical for phosphate compounds. The lead cation has seven oxygen neighbors with Pb--O distances ranging from 2.47~\r A to 2.99~\r A. The wide range of the Pb--O distances indicates the asymmetry of the local environment due to the presence of the $6s^2$ lone pair. Finally, zinc atoms form nearly regular tetrahedra with an additional oxygen atom at a longer distance of 2.38(1)~\r A. 

\begin{table}
\caption{\label{coordinates}
Atomic positions and isotropic atomic displacement parameters ($U_{\text{iso}}$, in units of $10^{-2}$~\r A$^2$) for PbZnVO(PO$_4)_2$. All the atoms occupy the general $8c$ position.
}
\begin{ruledtabular}
\begin{tabular}{cccccc}
  Position & $x$ & $y$ & $z$ & $U_{\text{iso}}$ \\\hline
  Pb & 0.17982(7) & 0.00359(10) & 0.41282(4) & 2.4(1) \\
  Zn & 0.1995(2) & 0.91225(16) & 0.08682(13) & 1.4(1) \\
  V & 0.0027(3) & 0.1605(3) & 0.21307(17) & 0.7(1) \\
  P(1) & 0.2529(4) & 0.4120(4) & 0.2407(2) & 0.6(1) \\
  P(2) & 0.4519(5) & 0.7997(5) & 0.5223(2) & 0.6(1) \\
  O(1) & 0.0664(9) & 0.6680(8) & 0.4137(5) & 0.3(1) \\
  O(2) & 0.1502(10) & 0.8214(9) & 0.2926(5) & 0.3(1) \\
  O(3) & 0.3087(12) & 0.9079(8) & 0.5224(4) & 0.3(1) \\
  O(4) & 0.0275(10) & 0.1531(9) & 0.2991(5) & 0.3(1) \\
  O(5) & 0.3932(9) & 0.6386(9) & 0.5341(4) & 0.3(1) \\
  O(6) & 0.3389(8) & 0.4965(14) & 0.1853(3) & 0.3(1) \\
  O(7) & 0.5217(11) & 0.8014(10) & 0.4456(5) & 0.3(1) \\
  O(8) & 0.1645(9) & 0.5077(12) & 0.2910(3) & 0.3(1) \\
  O(9) & 0.1528(11) & 0.3183(8) & 0.1870(4) & 0.3(1) \\
\end{tabular}
\end{ruledtabular}
\end{table}
  
Similar to the other AA$'$VO(PO$_4)_2$ compounds, there are two inequivalent phosphorous sites in the crystal structure. The P(1)O$_4$ tetrahedra share corners to the VO$_5$ pyramids and form [VOPO$_4$] layers. The P(2)O$_4$ tetrahedra along with the Pb and Zn cations reside in the interlayer space and can be considered as two-dimensional [AA$'$PO$_4$] blocks (Fig.~\ref{fig_structure}). The [VOPO$_4$] layers are buckled with the buckling angle $\varphi=163^{\circ}$ [compare to $160^{\circ}$ in BaZnVO(PO$_4)_2$ and $150^{\circ}$ in SrZnVO(PO$_4)_2$]. The VO$_5$ pyramids are connected via PO$_4$ tetrahedra that give rise to four inequivalent exchange couplings: $J_1,J_1'$ between nearest-neighbors and $J_2,J_2'$ between next-nearest-neighbors (left panel of Fig.~\ref{fig_structure}). Specific structural features of the magnetic layer along with the resulting exchange interactions constitute the main subject of the present paper and will be thoroughly discussed in the following sections. In the remainder of this section, we will briefly comment on the structure of the [AA$'$PO$_4$] blocks and on the role of the lead cation in PbZnVO(PO$_4)_2$ as compared to the other AA$'$VO(PO$_4)_2$ compounds.

\begin{table}
\caption{\label{distances}
Relevant interatomic distances (in~\r A) for the interactions $J_2$ and $J_2'$. Vanadium is surrounded by six oxygen atoms. However, two of these atoms occupy axial apices of the octahedron and do not take part in the superexchange. Phosphorous is coordinated by four oxygen atoms, while the two O--O distances measure the edges of the PO$_4$ tetrahedron.
}
\begin{ruledtabular}
\begin{tabular}{ccc@{\hspace{1.5cm}}ccc}
 & Distance & $J$ & & Distance & $J$ \\\hline
 V--O(2) & 2.00(1) & $J_2$  & P(1)--O(2) & 1.50(1) & $J_2$ \\
 V--O(8) & 2.03(1) & $J_2$  & P(1)--O(8) & 1.46(1) & $J_2$ \\
 V--O(6) & 2.13(1) & $J_2'$ & P(1)--O(6) & 1.47(1) & $J_2'$ \\
 V--O(9) & 2.00(1) & $J_2'$ & P(1)--O(9) & 1.58(1) & $J_2'$ \\
 V--O(4) & 1.55(1) &        & O(6)--O(9) & 2.27(1) & $J_2$ \\
 V--O(1) & 2.38(1) &        & O(2)--O(8) & 2.32(1) & $J_2'$ \\
\end{tabular}
\end{ruledtabular}
\end{table}

The crystal structures of the AA$'$VO(PO$_4)_2$ compounds contain A and A$'$ cations that reside in cavities formed by the PO$_4$ tetrahedra (right panel of Fig.~\ref{fig_structure}). The arrangement of these tetrahedra is rather flexible and can be tuned for a specific metal cation. In the previous study, we have shown that barium coordinates all the four tetrahedra around the cavity, while strontium prefers a lower coordination number and links to three tetrahedra only (see Fig.~6 in Ref.~\onlinecite{tsirlin2009}). Within a naive picture, the lead-containing structure should resemble the strontium-containing counterpart due to the similar ionic radii of Pb (1.23~\r A) and Sr (1.26~\r A).\cite{shannon} However, links to all the four tetrahedra around the cavity are present (right panel of Fig.~\ref{fig_structure}). This feature explains the close similarity of the PbZnVO(PO$_4)_2$ and BaZnVO(PO$_4)_2$ structures, despite the smaller size of the Pb cation compared to Ba. The different behavior of lead and strontium is a well-known feature caused by the asymmetric local environment of lead (compare, e.g., SrVO$_3$ and PbVO$_3$).\cite{shpanchenko2004} Yet the similarity between lead and barium is not common. 

We should also note that PbZnVO(PO$_4)_2$ belongs to the family of the AA$'$VO(PO$_4)_2$ phosphates with A~$\neq$~A$'$. The lead-zinc compound is rather different from the pure lead vanadylphosphate Pb$_2$VO(PO$_4)_2$ with a monoclinic structure.\cite{shpanchenko2006} This indicates that the accommodation of the two asymmetric lead cations requires an overall distortion of the crystal structure, while the single lead cation in PbZnVO(PO$_4)_2$ can be tolerated by simple tilts of the PO$_4$ tetrahedra in the [AA$'$PO$_4$] interlayer blocks. Indeed, the PO$_4$ tetrahedra in SrZnVO(PO$_4)_2$ and BaCdVO(PO$_4)_2$ have faces parallel to the [VOPO$_4$] layers, while this is not the case for the PbZnVO(PO$_4)_2$ structure (compare the middle panel of Fig.~\ref{fig_structure} with Fig.~2 in Ref.~\onlinecite{tsirlin2009}).

\section{Thermodynamic properties}
\label{experiment}
\subsection{Magnetic susceptibility}
\begin{figure*}
\includegraphics{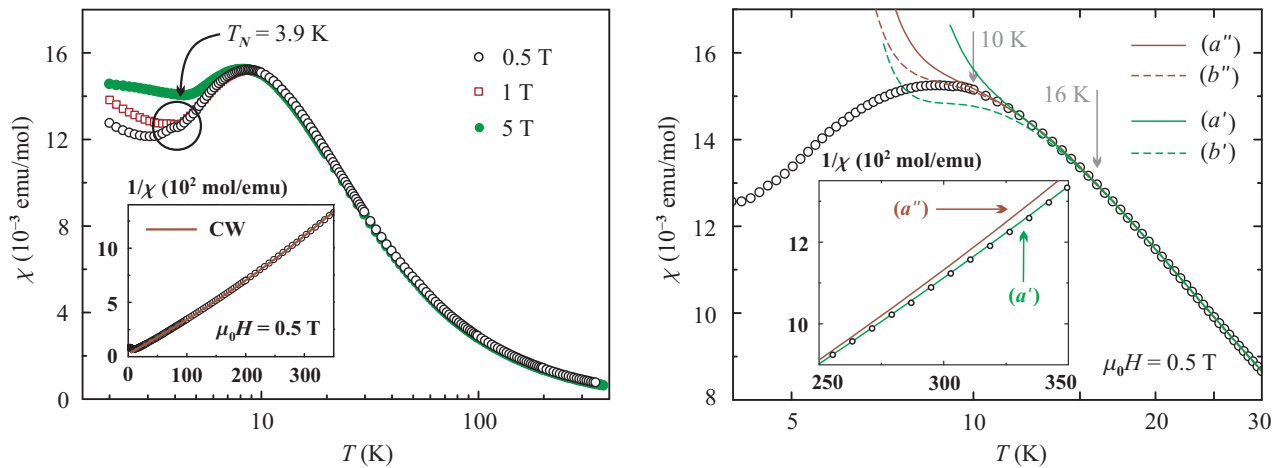}
\caption{\label{suscept}
(Color online) Temperature dependence of the magnetic susceptibility of PbZnVO(PO$_4)_2$. Left panel: curves measured at 0.5~T (open circles), 1~T (open squares), and 5~T (filled circles); the inset shows the Curie-Weiss (CW) fit above 40~K. Right panel: HTSE fits $(a')$ and $(a'')$ (solid lines), $(b')$ and $(b'')$ (dashed lines) according to Table~\ref{fits}; the inset shows the difference between $(a')$ and $(a'')$ in the high-temperature region, while dots indicate the experimental data.
}
\end{figure*}

Magnetic susceptibility ($\chi$) vs. temperature ($T$) data for PbZnVO(PO$_4)_2$ reveal a maximum at about 8.8~K (see Fig.~\ref{suscept}). Such maxima are typical for low-dimensional spin systems and indicate the onset of short-range spin correlations. Below the maximum, field-dependent behavior is observed. The data measured in low fields (below 1~T) show an anomaly at 3.9~K, while the data collected at 5~T present a bend at 4.2~K. These features are related to a phase transition that is also evidenced by a peak in the specific heat data (Fig.~\ref{fig_heat}). The field dependence of the transition temperature points to the magnetic origin of this transition. Based on the similarity to the other AA$'$VO(PO$_4)_2$ compounds, we can suggest that PbZnVO(PO$_4)_2$ undergoes antiferromagnetic ordering at $T_N=3.9$~K. The transition temperature is slightly increased in the magnetic field due to the suppression of quantum fluctuations. A similar effect has been recently observed in BaCdVO(PO$_4)_2$ (Ref.~\onlinecite{nath2008}) and in the unfrustrated square-lattice compound [Cu(HF$_2$)(pyz)$_2$]BF$_4$ (Ref.~\onlinecite{cupyz}). In the latter case, theoretical treatment was also given. Below the transition, the susceptibility remains nearly temperature-independent, as expected for a regular antiferromagnet.

The high-temperature part of the susceptibility curves shows a Curie-Weiss (CW) behavior. Above 40~K, we fit the data with the expression
\begin{equation}
  \chi=\chi_0^{\CW}+\dfrac{C}{T+\theta_{\CW}},
\end{equation}
where the temperature-independent term $\chi_0^{\CW}$ accounts for the diamagnetic and Van Vleck contributions, $C$ is the Curie constant, and $\theta_{\CW}$ is the Curie-Weiss temperature. The fit (see the inset in the left panel of Fig.~\ref{suscept}) resulted in $\chi_0^{\CW}=-1.8(1)\cdot 10^{-4}$~emu/mol, $C=0.354(1)$~emu~K/mol, and $\theta_{\CW}=6.8(1)$~K. The $C$ value corresponds to the effective magnetic moment of 1.64(1)~$\mu_B$ which is slightly reduced as compared to the expected spin-only value of 1.73~$\mu_B$ for V$^{+4}$. This reduction can be attributed to a slight reduction of the $g$-factor of vanadium due to the weak spin-orbit coupling, to the effect of the non-magnetic impurity phase, and to a very small amount of V$^{+5}$ defects in PbZnVO(PO$_4)_2$. 

To evaluate exchange couplings in PbZnVO(PO$_4)_2$, the susceptibility data are fitted with a high-temperature series expansion (HTSE) for the regular FSL model.\cite{HTSE} In Sec.~\ref{structure}, we have shown that the structure of PbZnVO(PO$_4)_2$ is rather complex, hence an extended spin model with four inequivalent couplings should be used. However, the spatial anisotropy (the difference between $J_1$ and $J_1'$ or $J_2$ and $J_2'$) has a minor effect on thermodynamic properties at sufficiently high temperatures, thus the expression for the regular model can be used. The resulting parameters should be considered as averaged couplings: $\bar J_1=(J_1+J_1')/2$ and $\bar J_2=(J_2+J_2')/2$.\cite{tsirlin2009}

The HTSE fits are known to be ambiguous. First, multiple solutions can be obtained, because individual couplings are separated in high orders of the expansion only, while the lowest, second-order term leads to a linear combination of the couplings (similar to the Curie-Weiss temperature $\theta_{\CW}$). Second, the precise $J$ values depend on the temperature range used. The temperature range of the fitting should lie within the convergence domain of the series, while the latter depends on the actual $J$ values and makes the fitting procedure iterative. The ambiguity due to the multiple solutions can be handled via a reference to independent experimental data (saturation field\cite{nath2008} or specific heat\cite{enrique}). Nevertheless, to remove the second ambiguity and to obtain accurate $J$ values, one has to vary the lower boundary of the fitting range ($T_{\min}$) and to check the convergence of the series at $T_{\min}$.

The experimental susceptibility data are readily fitted with the HTSE expression supplemented with an additional temperature-independent $\chi_0$ term. Similar to the other AA$'$VO(PO$_4)_2$ compounds,\cite{nath2008,enrique} we find two different solutions: $(a)$ $\bar J_1<0$, $\bar J_2>0$; and $(b)$ $\bar J_1>0$, $\bar J_2<0$. The $J$ values for both solutions at different $T_{\min}$ are listed in Table~\ref{fits}. To discuss the dependence on $T_{\min}$, we will use the solution $(a)$ as an example. 

\begin{figure}
\includegraphics{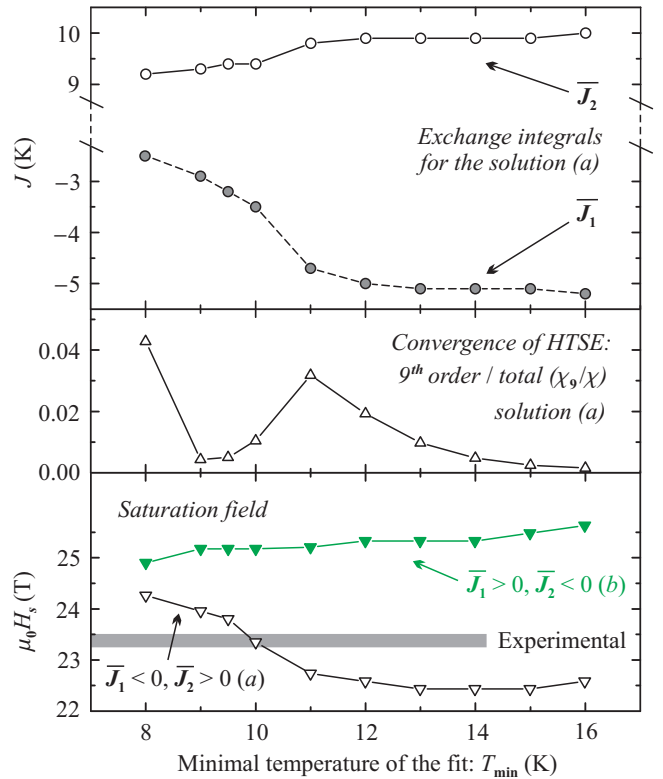}
\caption{\label{tmin}
(Color online) Results of the susceptibility fit depending on the minimum temperature of the fitting range ($T_{\min}$). Upper panel: averaged exchange couplings $\bar J_1$ (shaded circles) and $\bar J_2$ (open circles). Middle panel: convergence of the HTSE, measured as the ratio of the ninth-order term to the total susceptibility. Bottom panel: saturation fields for the solutions with $\bar J_1<0,\bar J_2>0$ (open triangles) and $\bar J_1>0,\bar J_2<0$ (filled green triangles); the shaded stripe shows the experimental value $\mu_0H_s\simeq 23.4$~T with the error bar of 0.2~T (see Ref.~\onlinecite{high-field} for details). The upper and middle panels refer \emph{only} to the solution $(a)$ with $\bar J_1<0,\bar J_2>0$, while the bottom panel shows \emph{both} the solutions.
}
\end{figure}

The value of $T_{\min}$ is varied between 8~K and 16~K. The upper panel of Fig.~\ref{tmin} shows that the $\bar J$ values are rather stable for $T_{\min}=11-16$~K. However, lower $T_{\min}$ leads to a slight decrease of $\bar J_2$ and to a notable increase of $\bar J_1$, thus reducing the effective frustration ratio \mbox{$\alpha=\bar J_2/\bar J_1$}. To check the convergence of the series at $T_{\min}$, we use the ratio of the highest-order HTSE term to the susceptibility value. The change in the convergence is non-monotonous (see the middle panel of Fig.~\ref{tmin}). As $T_{\min}$ is decreased from 16~K to 11~K, the convergence is deteriorated. Yet the change in the frustration ratio for the solutions at low $T_{\min}$ improves the convergence for $T_{\min}=9-10$~K. Basically, we find two equally reliable solutions that correspond to $T_{\min}=16$~K $(a')$ and $T_{\min}=9-10$~K $(a'')$. 

\begin{table*}
\caption{\label{fits}
HTSE fits of the magnetic susceptibility data. $T_{\min}$ is the minimum temperature of the fitting range, $\chi_0$ is the temperature-independent contribution, $g$ is the $g$-factor, $\bar J_1$ and $\bar J_2$ are averaged NN and NNN couplings, respectively, and $\mu_0H_s^{\reg}$ is the saturation field for the regular FSL model, as calculated via Eq.~\eqref{saturation}. $(a)$ and $(b)$ refer to the solutions with different signs of $\bar J_1$ and $\bar J_2$. The primes discriminate the solutions at different $T_{\min}$.
}
\begin{ruledtabular}
\begin{tabular}{ccccccc}
  Fit   & $T_{\min}$ (K) & $\chi_0$ (10$^{-4}$~emu/mol) & $g$     & $\bar J_1$ (K) & $\bar J_2$ (K) & $\mu_0H_s^{\reg}$ (T) \\\hline
  $(a')$ & 16  & $-1.71(3)$  & 1.865(2) & $-5.2(2)$      & $10.0(1)$  &  $22.6$    \\
 $(a'')$ & 10  & $-2.11(4)$  & 1.889(2) & $-3.5(1)$      & $9.4(1)$   &  $23.4$    \\
  $(b')$ & 16  & $-1.84(3)$  & 1.877(2) & $8.4(1)$       & $-2.6(1)$  &  $25.6$    \\
 $(b'')$ & 10  & $-2.08(4)$  & 1.892(2) & $8.3(1)$       & $-1.7(1)$  &  $25.2$    \\
\end{tabular}
\end{ruledtabular}
\end{table*}

Now, we will select the correct solution by calculating the saturation field and comparing it to the experimental value of $\mu_0H_s=23.4$~T.\cite{high-field} At $J_2/|J_1|>0.5$ [solution $(a)$], the saturation field of the \emph{regular} FSL model is expressed as follows:\cite{thalmeier2008}
\begin{equation}
  \mu_0H_s^{\reg}=2k_B(\bar J_1+2\bar J_2)/(g\mu_B),
\label{saturation}\end{equation}
where the superscript "$\reg$" denotes the saturation field for the regular model, and $g$ is the $g$-factor (we use $g=1.95$ as a representative value for V$^{+4}$-containing compounds\cite{sr2v3o9-esr}). Solution $(b)$ corresponds to the N\'eel ordering, and the saturation field depends on $\bar J_1$ only: $\mu_0H_s^{\reg}=4k_B\bar J_1/(g\mu_B)$. The resulting $\mu_0H_s^{\reg}$ values are plotted in the bottom panel of Fig.~\ref{tmin}. The solutions of type $(b)$ overestimate the saturation field by $1.5-2.0$~T and can be rejected, as further supported by band structure results (see Sec.~\ref{band}) suggesting the FM $\bar J_1$ and AFM $\bar J_2$. The solution $(a')$ underestimates the saturation field, while the solution $(a'')$ shows the best agreement with the experimental $\mu_0H_s$ of 23.4~T. At first glance, this result suggests $(a'')$ as the final answer. However, one should be aware of the difference between the regular model assumed in \eqref{saturation} and the real spin model with inequivalent couplings $J_1$ and $J_1'$ (Fig.~\ref{fig_structure}). In our previous work,\cite{high-field} we have shown that the difference between $J_1$ and $J_1'$ increases $\mu_0H_s$ compared to $\mu_0H_s^{\reg}$. Thus, both the solutions $(a')$ and $(a'')$ are possible and would correspond to different regimes of the extended FSL model in PbZnVO(PO$_4)_2$: $(a')$ fits better to the square lattice with spatial anisotropy of the NN couplings, while $(a'')$ corresponds to the regular square lattice with $J_1=J_1'$.

The final choice between $(a')$ and $(a'')$ can be made after a close examination of both fits. In the case of fit $(a'')$, the low-temperature part of the data is fitted at the cost of a less accurate high-temperature behavior. Plotting the $1/\chi$ curves in the high-temperature region, we find that the $(a'')$ fit slightly deviates from the experimental data, while the fit $(a')$ perfectly matches the data (see the inset in the right panel of Fig.~\ref{tmin}). This can also be seen from the $\chi_0$ values in Table~\ref{fits} compared to the Curie-Weiss fit value $\chi_0^{\CW}\simeq -1.8\cdot 10^{-4}$~emu/mol. The fit $(a'')$ overestimates $\chi_0$, hence the high-temperature behavior of the experimental data cannot be accurately reproduced. 

The above arguments provide strong evidence for the solution $(a')$ and for the scenario of the spatially anisotropic FSL. We further checked this conclusion by fitting the data collected in different fields. The $\bar J$ values from different data sets coincided within 0.1~K, i.e., within the error bar of the fit. Thus, the analysis of the susceptibility data provides accurate estimates of $\bar J_1$ and $\bar J_2$, while the further reference to the saturation field points to the spatial anisotropy of the NN couplings. According to Ref.~\onlinecite{high-field}, the difference between $\mu_0H_s$ and $\mu_0H_s^{\reg}$ is a measure of the spatial anisotropy: we find that $J_1'-J_1\simeq 1.1$~K. This value will be further compared with computational results in Sec.~\ref{band}. The saturation field could also be increased due to the AFM interlayer couplings.\cite{goddard2008} However, our computational estimates suggest that the interlayer couplings are below 0.1~K (see Sec.~\ref{band}), hence their effect on the saturation field value will not exceed the experimental error bar of $H_s$.

Our analysis demonstrates that HTSE fits of the susceptibility data yield accurate estimates of averaged exchange couplings on the square lattice. Due to the symmetry of the model hamiltonian, the fitting procedure is ambiguous. However, the reference to the experimentally measured saturation field can remove this ambiguity. Furthermore, it is crucial to collect the susceptibility data at sufficiently high temperatures, because the high-temperature limit constrains additional variable parameters, such as $\chi_0$ and $g$. In the following, we will show that the specific heat data can also be a key for the assignment of the $\bar J_1$ and $\bar J_2$ values, although such data are less sensitive to weak changes in the exchange couplings.

\subsection{Specific heat}
\begin{figure}
\includegraphics{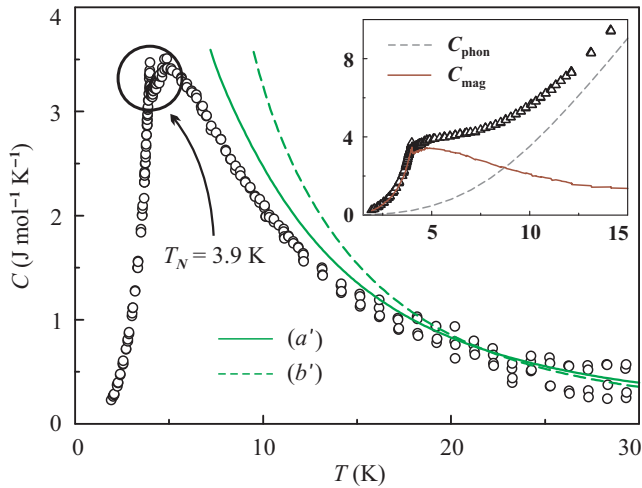}
\caption{\label{fig_heat}
(Color online) Specific heat of PbZnVO(PO$_4)_2$: magnetic contribution (open circles) and HTSE predictions for different solutions according to Table~\ref{fits}. The inset shows the experimental data (open triangles) along with the phonon (dashed line) and magnetic (solid line) contributions.
}
\end{figure}

The temperature dependence of the specific heat ($C_p$) for PbZnVO(PO$_4)_2$ is shown in the inset of Fig.~\ref{fig_heat}. At low temperatures, $C_p$ increases and has a peak at $T_N=3.9$~K. This peak matches the bends of the susceptibility curves (Fig.~\ref{suscept}) and indicates a magnetic phase transition. Above $T_N$, specific heat increases and shows a pronounced bend at 7.5~K. This region can be interpreted as a sum of the rapidly increasing phonon contribution ($C_{\ph}$) and the magnetic contribution ($C_{\mg}$) that shows a maximum due to the correlated spin excitations right above $T_N$. Since the exchange couplings in PbZnVO(PO$_4)_2$ are relatively weak, one can expect $C_{\mg}$ to reach its high-temperature limit ($\sim 1/T^2$) above $18-20$~K. Then, the data can be fitted with the expression
\begin{equation}
 C_{p}(T)=\frac{A}{T^{2}}+9R\sum_{n=1}^{n=5}c_{n}\left( \frac{T}{\theta_D^{(n)}} \right)^{3}\int\limits_{0}^{\theta_D^{(n)}/T}\frac{x^{4}e^{x}}{\left(  e^{x}-1\right)^2}\,dx, \label{Debye}%
\end{equation}
where the first term describes the high-temperature part of $C_{\mg}$, while further terms are Debye functions with Debye temperatures $\theta_D^{(n)}$ and integer coefficients $c_n$. A similar model for $C_{\ph}$ has been used in our previous studies of vanadium phosphates.\cite{nath2008,enrique,kini2006,nath2008-2} 

Equation \eqref{Debye} leads to a perfect fit of the experimental data above 18~K. The resulting $A=353(41)$~J~K/mol has a large error bar, because at high temperatures $C_{\mg}$ gives a small contribution to the total specific heat. The $A$ value is in reasonable agreement with the $\bar J$ values: according to Ref.~\onlinecite{HTSE}, $A/T^2$ is the leading term in the HTSE for the specific heat, and
\begin{equation}
  A=(3R/8)(\bar J_1^2+\bar J_2^2).
\label{heat}\end{equation}  
The solution $(a')$ yields $A\simeq 396$~J~K/mol, while the solutions $(b')$ and $(b'')$ lead to $A=220-240$~J~K/mol which is well below the experimental value. However, the distinction between the solutions $(a')$ and $(a'')$ is hardly possible, because both the respective $A$ values (396 and 314~J~K/mol) calculated from Eq.~\eqref{heat} agree reasonably well to the fitted value of 353~J~K/mol.

The $\theta_D^{(n)}$ values are further used to estimate $C_{\ph}$ in the low-temperature region and to subtract the phonon contribution from the total specific heat. The resulting magnetic contribution is shown in Fig.~\ref{fig_heat}. To test the correctness of the procedure, we estimate magnetic entropy $S$ by integrating the $C_{\mg}/T$ curve. We obtain $S\simeq 5.04$~J/mol~K, i.e., the magnetic entropy is slightly underestimated with respect to the expected value of $R\ln 2\simeq 5.76$~J/mol~K. This underestimate amounts to 12~\%\ and can be ascribed to a slight overestimate of the low-temperature phonon contribution and to a small amount of an impurity phase.

The extracted experimental $C_{\mg}(T)$ curve can be compared to the HTSE predictions.\cite{HTSE} At high temperatures, the error bar for $C_{\mg}$ is very high, because the measured heat capacity is dominated by the phonon contribution. Therefore, the direct fit of the $C_{\mg}(T)$ curve is impossible. Nevertheless, one can use the intermediate region ($10-20$~K) and compare the data with the results of the susceptibility fitting. In Fig.~\ref{fig_heat}, we plot the HTSE for the solutions $(a')$ and $(b')$ (Table~\ref{fits}). The curve for the solution $(a')$ is in reasonable agreement with the experimental data (the agreement is further improved if the entropy weight error of about 10~\% is corrected). The solution $(b')$ clearly overestimates $C_{\mg}$ below 15~K. Although both the solutions are obtained by fitting the susceptibility data \emph{above} 16~K, their convergence domains extend to lower temperatures (see the middle panel of Fig.~\ref{tmin} for the convergence of the susceptibility series). Therefore, the comparison in the region around 15~K is still reasonable. Due to the limited temperature range of the applicable specific heat data, distinguishing between $(a')$ and $(a'')$ is hardly possible.

Below $T_N$, the specific heat of PbZnVO(PO$_4)_2$ rapidly decreases. Such a behavior is qualitatively consistent with antiferromagnetic ordering. However, the temperature dependence of $C_{\mg}$ in the ordered phase does not follow a simple $T^3$ trend expected for a regular antiferromagnet. The origin of this behavior is presently unclear. Similar deviations from the $T^3$ behavior were observed in the other FSL compounds.\cite{enrique} In these compounds, the columnar AFM ordering is stabilized by quantum fluctuations that select the collinear ground state among numerous ground states, which are degenerate in the classical Heisenberg model (the so-called order-from-disorder mechanism).\cite{shannon2004} One might suggest that this feature modifies the magnetic excitation spectrum and changes the low-temperature behavior of the specific heat. To clarify this issue, further experimental and theoretical studies are desirable.
\medskip

To conclude the experimental section, the magnetic susceptibility and the specific heat data for PbZnVO(PO$_4)_2$ can be interpreted within the FSL model. Thermodynamic properties consistently point to ferromagnetic NN and antiferromagnetic NNN couplings in PbZnVO(PO$_4)_2$ with the accurate estimates of $\bar J_1$ and $\bar J_2$. The reference to the saturation field suggests the spatial anisotropy of the NN couplings. In the following sections, we will use a microscopic approach in order to get further insight into the spin system of PbZnVO(PO$_4)_2$ and to elucidate the relation between the structural features of this compound and individual exchange couplings.

\section{Evaluation of the exchange integrals}
\label{band}
The calculated band structure of the material can be used to evaluate individual exchange couplings and to construct a microscopic model of magnetic interactions. In the case of magnetic insulators, there are two well-known approaches for the evaluation of the exchange couplings. The first approach utilizes the uncorrelated (LDA) band structure which is further mapped onto a TB model and, subsequently, onto a Hubbard model in the strongly correlated regime. Then, at half-filling the low-lying excitations of the system are properly described within a Heisenberg model, while the parameters of this model are expressed via microscopic parameters of the starting Hubbard model. The alternative approach is based on a correlated band structure which is usually obtained via local spin density approximation (LSDA)+$U$ calculations that treat correlations in a mean-field way. To estimate exchange couplings, one can calculate total energies for ordered spin configurations and map the resulting energies onto the classical Heisenberg model. However, alternative methods, utilizing Green's functions\cite{lichtenstein} and a more sophisticated treatment of electronic correlations,\cite{savrasov} are also possible. 

The key feature of the second approach is the sizable and implicit dependence of the resulting exchange couplings on the parameters describing electronic correlations -- in particular, on the on-site Coulomb repulsion parameter $U_d$. While in many systems the exchange couplings are sufficiently large and the uncertainty in $U_d$ leads to $10-20$~\% uncertainty in $J$, the materials with weak exchange couplings are more difficult to study. In previous works,\cite{tsirlin2009,nath2008-2} we have shown that a change of 1~eV in $U_d$ can modify the signs of the exchange coupling constants in vanadium phosphates, thus different magnetic ground states are obtained. A careful fitting of the $U_d$ value to the experimental data (e.g., the Curie-Weiss temperature) helps to establish the correct scenario. Still, it is preferrable to use the model approach and to keep the explicit dependence on the Coulomb repulsion parameter, while studying materials with weak exchange couplings. Another advantage of the model approach is the insight into the microscopic mechanism of the exchange coupling, because contributions of different orbitals are easily separated.

\begin{figure}
\includegraphics{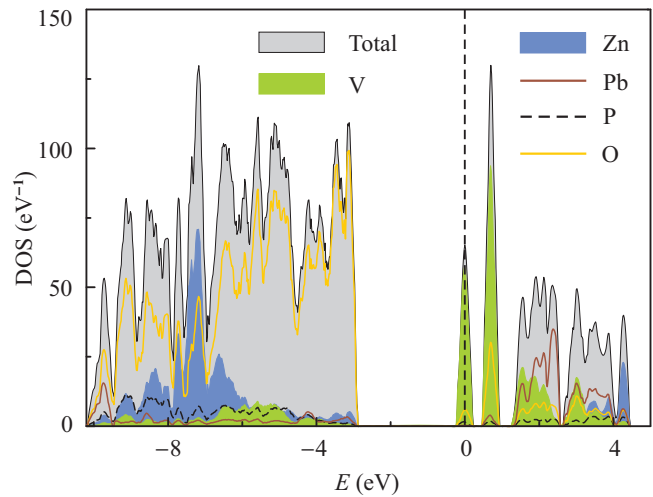}
\caption{\label{dos}
(Color online) LDA density of states plot for PbZnVO(PO$_4)_2$: filled regions denote the total DOS, V and Zn contributions. The contributions of Pb, O, and P are shown by dark solid, light solid, and dashed lines, respectively. The Fermi level is at zero energy.
}
\end{figure}

\begin{figure}
\includegraphics{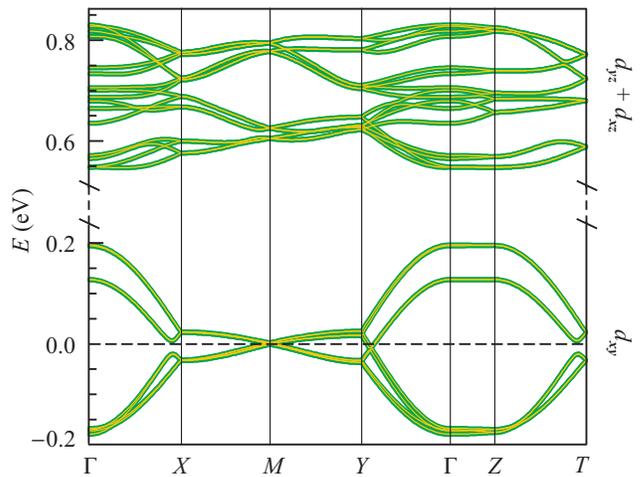}
\caption{\label{bands}
(Color online) LDA band structure of PbZnVO(PO$_4)_2$ (thin light lines) and the fit of the three-orbital TB model (thick lines). The Fermi level is at zero energy. The notation of $k$ points is as follows: $\Gamma(0,0,0)$, $X(0.5,0,0)$, $M(0.5,0.5,0)$, $Y(0,0.5,0)$, $Z(0,0,0.5)$, and $T(0.5,0,0)$, where the coordinates are given in units of the respective reciprocal lattice parameters. Note that most of the bands are close to double degeneracy due to the weak interlayer coupling.
}
\end{figure}

Following the model approach, we consider the uncorrelated (LDA) band structure. The calculated density of states (DOS) is shown in Fig.~\ref{dos}. The valence bands of PbZnVO(PO$_4)_2$ are formed by oxygen $2p$ orbitals with sizable contributions from Pb $6s$ states at $-10$~eV and from Zn $3d$ states at about $-7$~eV. The narrow bands at the Fermi level have vanadium $3d_{xy}$ origin, while the other isolated band complex at $0.5-0.9$~eV is formed by vanadium $3d_{xz}$ and $3d_{yz}$ states (see also Fig.~\ref{bands}). The $e_g$ states of vanadium lie above 1~eV and show strong hybridization with the $6p$ states of lead. The obtained crystal field splitting of the vanadium orbitals is consistent with the square-pyramidal (or distorted-octahedral) environment. Similar splittings have been observed in other vanadium compounds.\cite{tsirlin2009,tsirlin2008-2}

The minimal microscopic model should include 8 $d_{xy}$ bands lying near the Fermi level (Fig.~\ref{bands}). These bands originate from 8 vanadium atoms in the unit cell and are close to double degeneracy due to the weak interlayer couplings. The $d_{xy}$ bands are fitted with a one-orbital TB model. The hoppings $t^{xy\rightarrow xy}$ of this TB model (Table~\ref{exchanges}) are further introduced into a one-band Hubbard model with the effective on-site Coulomb repulsion $U_{\eff}$. The $U_{\eff}$ parameter is of the order of several~eV.\cite{tsirlin2009,tsirlin2008-2} Then, the \mbox{$t\ll U_{\eff}$} condition and the half-filling regime justify the mapping onto the Heisenberg model for low-lying excitations, and AFM contributions to the exchange integrals are estimated via the well-known expression $J_{ij}^{\AFM}=4t_{ij}^2/U_{\eff}$.

To account for FM contributions to the exchange integrals, one has to extend the model by the inclusion of unoccupied states. The underlying physical reason is as follows. If an electron hops to a half-filled orbital, its spin should be antiparallel to the spin of the electron on the destination site, hence AFM coupling is realized. In contrast, the hopping to an empty orbital favors parallel arrangement of spins due to the Hund's coupling (on-site exchange) on the destination site. Analytical expressions for the FM coupling of this type were first derived by Kugel and Khomskii as a special case of their model.\cite{kugel1982} Later on, this expression was employed by Mazurenko \textit{et al.}\cite{mazurenko2006} for the analysis of the exchange couplings in Na$_2$V$_3$O$_7$. Following Refs.~\onlinecite{kugel1982} and~\onlinecite{mazurenko2006}, we estimate the FM contributions to the exchange couplings as
\begin{equation}
  J_{ij}^{\FM}=-\sum_{\alpha}\dfrac{(t_{ij}^{xy\rightarrow\alpha})^2} {(U_{\eff}+\Delta_{\alpha})(U_{\eff}-J_{\eff}+\Delta_{\alpha})},
\label{jfm}\end{equation}
where the index $\alpha$ indicates empty orbitals, $t_{ij}^{xy\rightarrow\alpha}$ denotes the hopping from the $d_{xy}$ orbital on site $i$ to an $\alpha$ orbital on site $j$,\cite{note1} $\Delta_{\alpha}=\eps_{\alpha}-\eps_{xy}$ is the energy splitting between the occupied ($xy$) and empty ($\alpha$) orbitals, $U_{\eff}$ and $J_{\eff}$ are the effective on-site Coulomb repulsion and the on-site (Hund's) exchange parameters, respectively. Hereby, we assume that the on-site repulsion and exchange are similar for the $xy$ and for all the $\alpha$ orbitals.

\begin{table}
\caption{\label{exchanges}
Parameters of the TB model and the resulting exchange couplings in PbZnVO(PO$_4)_2$. $t^{xy\rightarrow xy}$ denote the hoppings between $d_{xy}$ orbitals, $J^{\AFM}$ and $J^{\FM}$ are antiferromagnetic and ferromagnetic contributions to the total exchange $J$, respectively.
}
\begin{ruledtabular}
\begin{tabular}{ccccc}
  & $t^{xy\rightarrow xy}$ (meV) & $J^{\AFM}$ (K) & $J^{\FM}$ (K) & $J$ (K) \\
  $J_1$  & 11    & 1.3  & $-4.6$ & $-3.3$ \\
  $J_1'$ & 14    & 2.0  & $-4.2$ & $-2.2$ \\
  $J_2$  & $-46$ & 21.9 & $-0.1$ & $21.8$ \\
  $J_2'$ & $-38$ & 15.0 & $-0.1$ & $14.9$ \\
\end{tabular}
\end{ruledtabular}
\end{table}

The hoppings $t^{xy\rightarrow\alpha}$ are estimated from a fit of the LDA band structure with a multi-orbital TB model. For the evaluation of $J^{\FM}$, we include in the model both the $d_{yz}$ and $d_{xz}$ states of vanadium (see Fig.~\ref{bands}) and omit the $e_g$ states due to their strong hybridization with lead orbitals. This approach is justified by a reference calculation for the isostructural SrZnVO(PO$_4)_2$ (not shown), where the $e_g$ states are weakly hybridized and show small hoppings to the $xy$ states ($t^{xy\rightarrow e_g}<5$~meV). Another justification is given by an extended TB model including all the vanadium $3d$ and, additionally, lead $6p$ states. Due to the presence of uncorrelated (Pb $6p$) states, this extended model can not be directly mapped onto the Heisenberg model, and one has to reduce the model to five vanadium orbitals first. Still, the parameters of the extended model suggest weak hoppings between the half-filled $d_{xy}$ orbitals and the $e_g$ orbitals of vanadium or the $6p$ orbitals of lead ($t^{xy\rightarrow e_g},t^{xy\rightarrow \text{Pb6}p}<2$~meV). Therefore, the $xy\rightarrow e_g$ hoppings in the reduced (five-orbital) model should also be weak. 

For the $U_{\eff}$ and $J_{\eff}$ parameters, we use the representative values of 4.5~eV and 1~eV, as employed in the previous model analysis.\cite{tsirlin2009,nath2008-2,tsirlin2008-2,mazurenko2006} We should note that the $J_{\eff}$ value has little influence on $J^{\FM}$, and the main uncertainty arises from $U_{\eff}$ ($J^{\FM}$ roughly scales as $1/U_{\eff}^2$). However, the change of $U_{\eff}$ in the reasonable range of $4-6$~eV leads to a moderate change in the resulting $J$ values, despite all the couplings are weak. Such a moderate dependence on $U_{\eff}$ is a clear advantage of the model approach as compared to the LSDA+$U$ calculations (see Ref.~\onlinecite{tsirlin2009}).

The calculated exchange integrals are listed in Table~\ref{exchanges} along with their FM and AFM contributions. We find that the NN couplings $J_1$ and $J_1'$ are FM, while the NNN couplings $J_2$ and $J_2'$ are AFM in remarkable agreement with the experimental results: $\bar J_1\simeq -5.2$~K and \mbox{$\bar J_2\simeq 10.0$~K}. Moreover, even the difference $J_1'-J_1\simeq 1.1$~K perfectly matches the experimental value of 1.1~K. The interactions $J_1$ and $J_1'$ show both FM and AFM contributions, although the former contribution dominates. In the case of $J_2$ and $J_2'$, the values are somewhat overestimated as compared to $\bar J_2\simeq 10$~K. This effect is also observed for the other AA$'$VO(PO$_4)_2$ vanadium phosphates.\cite{tsirlin2009} The NNN interactions show negligible FM contributions consistent with long V--V separations of about 6.5~\r A. The largest interlayer coupling is of the order of 0.01~K.

Apart from the support of the experimental results, our microscopic model provides further insight into the spin system of PbZnVO(PO$_4)_2$. In particular, the computational study gives independent estimates of $J_2$ and $J_2'$. We find pronounced spatial anisotropy of the NNN couplings ($J_2'/J_2\simeq 0.68$) which is comparable to that in BaZnVO(PO$_4)_2$ ($J_2'/J_2\simeq 0.61$) and Pb$_2$VO(PO$_4)_2$ ($J_2'/J_2\simeq 0.67$).\cite{tsirlin2009} In Ref.~\onlinecite{tsirlin2009}, we have shown that the spatial anisotropy of the NNN couplings is mainly determined by the difference in V--O distances for the respective superexchange pathways. In the next section, we will discuss in more detail the geometry of the superexchange pathways in PbZnVO(PO$_4)_2$ and the relation between the structural features and magnetic interactions in FSL compounds.

\section{Modeling}
\label{modeling}
\begin{figure*}
\includegraphics{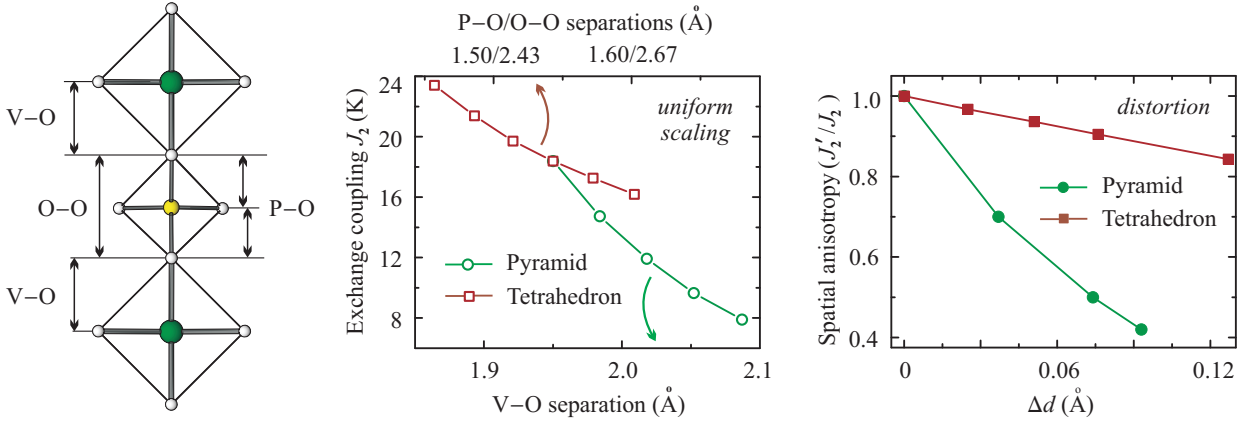}
\caption{\label{fig_model}
(Color online) The superexchange pathway comprising the single tetrahedral PO$_4$ group (left panel) and the evolution of the exchange couplings upon variation of the geometry of this pathway (middle and right panels). The middle panel shows the change in the exchange coupling $J_2$ along the scaling of the polyhedra in a uniform way, while the right panel presents the spatial anisotropy of the exchange couplings ($J_2'/J_2$) due to the different P--O or V--O distances in the respective polyhedra ($\Delta d$ is the difference between the distances, relevant for $J_2$ and $J_2'$).\cite{note3} While the parameters of one polyhedron (pyramid or tetrahedron) are varied, the geometry of the other polyhedron is fixed at $d\text{(V--O)}=1.95$~\r A for the pyramid and $d\text{(P--O)}=1.54$~\r A for the tetrahedron (the crossing point of the curves in the middle panel).
}
\end{figure*}

The magnitudes of superexchange interactions depend delicately on geometrical parameters of the superexchange pathway. In the conventional M--O--M superexchange scenario (M is a transition metal), the exchange integral is most sensitive to the angle at the oxygen atom.\cite{goodenough} However, more complicated superexchange pathways can show different trends. In the case of vanadium phosphates (and, more generally, any transition-metal compounds with XO$_4$ tetrahedra, X being a main-group cation), the typical scenario is the ``superexchange via a single tetrahedron'', as shown in the left panel of Fig.~\ref{fig_model}. This scenario corresponds to the interactions $J_2$ and $J_2'$ in the layered vanadium phosphates AA$'$VO(PO$_4)_2$. According to Ref.~\onlinecite{tsirlin2009}, the A and A$'$ metal cations have influence on the: (i) buckling of the [VOPO$_4$] layer, i.e., on the angles at oxygen atoms that determine the orientation of the tetrahedron with respect to the VO$_5$ pyramid; (ii) stretching of the [VOPO$_4$] layer, i.e., vanadium--oxygen distances. Our microscopic analysis has shown that the exchange integrals are more sensitive to the latter, while the change in the angles (in a reasonable range) leads to a weak modification of the exchange couplings (see Fig.~8 in Ref.~\onlinecite{tsirlin2009}). These findings suggest that the O--(P)--O interaction is rather insensitive to the mutual orientation of the VO$_5$ pyramids and the PO$_4$ tetrahedron. However, the microscopic origin of this interaction remains unclear.

We start with the careful analysis of individual structural parameters of PbZnVO(PO$_4)_2$. To characterize the $J_2$ and $J_2'$ superexchange pathways, we average the respective V--O distances (Table~\ref{distances}): $d\simeq 2.01$~\r A ($J_2$), $d'\simeq 2.05$~\r A ($J_2'$). According to Ref.~\onlinecite{tsirlin2009}, these numbers mainly determine the magnitudes of $J_2$ and $J_2'$. The $d$ value is somewhat higher as compared to the other AA$'$VO(PO$_4)_2$ compounds [e.g., $d\simeq 1.98$~\r A in SrZnVO(PO$_4)_2$]. In consequence, the interaction $J_2$ in PbZnVO(PO$_4)_2$ should be weaker than in the SrZn compound. However, this is not the case: we find $J_2\simeq 22$~K (Table~\ref{exchanges}) in contrast to $J_2^{\AFM}\simeq 19$~K in SrZnVO(PO$_4)_2$.\cite{tsirlin2009} Thus, other structural factors influence on the $J_2$ value. Indeed, several P--O distances in PbZnVO(PO$_4)_2$ are smaller than in the other compounds of the family. This gives the clue that the magnitude of the superexchange interaction is also controlled by the structural parameters of the PO$_4$ tetrahedron.

To analyze the issue in more detail, we perform calculations for model structures. Such structures include the magnetic [VOPO$_4$] layers separated by Li cations. The details of the procedure can be found in Ref.~\onlinecite{tsirlin2009}. We assume that the magnetic interactions are determined by the structure of the [VOPO$_4$] layer, while metallic cations simply provide proper charges (then, the complex [AA$'$PO$_4$]$^{+1}$ block can be substituted by a layer of Li$^{+1}$ cations). This assumption perfectly holds for the AFM interactions (i.e., for the $xy\rightarrow xy$ hoppings),\cite{tsirlin2009} while the case of the FM interactions is more problematic (hoppings between different $d$ orbitals depend on the structure of interlayer blocks). In the following, we focus on the AFM interactions ($J_2$ and $J_2'$) only. Clearly, the FM contributions to these interactions are negligible due to the large V--V separation (see Table~\ref{exchanges}), hence $J_2\simeq J_2^{\AFM}$. 

First, we consider the uniform scaling of individual polyhedra in the magnetic layer and trace the change in the coupling constant $J_2$ (due to the uniform scaling, the layer remains regular, hence $J_2=J_2'$). Both the pyramid and the tetrahedron exhibit sizable influence on the exchange integral (see middle panel of Fig.~\ref{fig_model}). The trend for the pyramid basically reproduces the change in $J_2'$ in the AA$'$VO(PO$_4)_2$ compounds: $d'$ is increased from 2.00~\r A in BaCdVO(PO$_4)_2$ up to 2.08~\r A in SrZnVO(PO$_4)_2$; $J_2'$ is correspondingly reduced from 17~K down to 7~K.\cite{tsirlin2009} The curve for the tetrahedron shows a smaller slope. Thus, the exchange coupling is less sensitive to the intermediate O--(P)--O link, and the V--O distances (i.e., the overlap between vanadium and oxygen orbitals) are more important.

Now, we will elucidate the relevant geometrical parameter of the tetrahedral group. Uniform scaling of the tetrahedron implies that both the P--O distances and the O--O distances (edges of the tetrahedron) are increased. In the next run of model calculations, we kept the O--O distances fixed and varied the P--O distances by shifting the phosphorous atom inside the tetrahedron. The shift of the P atom leads to the non-equivalence of $J_2$ and $J_2'$. However, the $J_2'/J_2$ remains close to unity even if the difference between the respective P--O distances is pronounced (see the right panel of Fig.~\ref{fig_model}). In contrast, the change in individual V--O distances leads to the rapid decrease of $J_2'/J_2$. This result enables us to establish the V--O distances in the pyramid and the O--O distances in the tetrahedron as the relevant geometrical parameters for the superexchange interactions $J_2$ and $J_2'$. 

\begin{figure}
\includegraphics{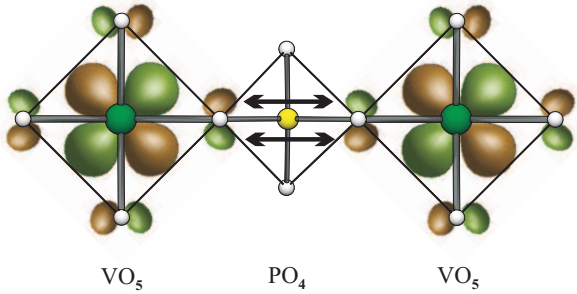}
\caption{\label{fig_wannier}
(Color online) Plot of the Wannier functions for the superexchange pathway, including two VO$_5$ pyramids and the single PO$_4$ tetrahedron. Arrows indicate the overlap of oxygen $p$ orbitals.
}
\end{figure}

A more general result of the above analysis deals with the contribution of the phosphorous atom to the superexchange. Since the P--O distances have little influence on the exchange integrals, we can conclude that the phosphorous orbitals play a minor role in the coupling. The superexchange is the result of the interaction between the oxygen atoms on the edge of the tetrahedron. This picture is illustrated by the plot of the Wannier functions in Fig.~\ref{fig_wannier}. Each Wannier function is composed of the vanadium $d_{xy}$ orbital and oxygen $p$ orbitals. As expected, the contribution of phosphorous orbitals is minor. Then, the superexchange can be considered as the $\pi$-overlap of vanadium and oxygen orbitals with a further $\pi$-overlap of the two oxygen orbitals on neighboring atoms. This scenario also accounts for different changes in $J_2$ upon scaling the pyramids and the tetrahedra (see middle panel of Fig.~\ref{fig_model}). The scaling of the pyramids modifies two V--O distances and has stronger influence on the exchange as compared to the scaling of the tetrahedra that leads to the change in a single O--O distance. In the next section, we will discuss the implications of this result for a tuning of the magnetic interactions in FSL materials.

\section{Discussion and summary}
\label{discussion}
Our studies identified PbZnVO(PO$_4)_2$ as a FSL material with ferromagnetic NN interactions, antiferromagnetic NNN interactions, and a spatial anisotropy of the exchange couplings. Thermodynamic measurements and band structure calculations provide convincing evidence for this scenario and establish solid and accurate estimates of individual exchange couplings. Now, one can compare PbZnVO(PO$_4)_2$ to other FSL materials and obtain deeper experimental insight into the properties of the spatially anisotropic FSL model.

\begin{table*}
\caption{\label{comparison}
Basic characteristics of thermodynamic data for the FSL compounds with FM $\bar J_1$ and AFM $\bar J_2$. $\alpha=\bar J_2/\bar J_1$ is the effective frustration ratio, $T_{\max}^{\,\chi}$ and $T_{\max}^{\,C}$ are the positions of the magnetic susceptibility and the specific heat maxima, while $\chi^{\max}$ and $C^{\max}$ are the absolute values at the maxima. $J_c=(\bar J_1^2+\bar J_2^2)^{1/2}$ defines the thermodynamic energy scale.
}
\begin{ruledtabular}
\begin{tabular}{cc@{\hspace{1.5cm}}cccc@{\hspace{1.5cm}}c}
  Compound & $\alpha$ & $T_{\max}^{\,\chi}/J_c$ & $\chi^{\max}J_c$ & $T_{\max}^{\,C}/J_c$ & $C^{\max}/R$ & Ref. \\
  PbZnVO(PO$_4)_2$   & $-1.9$ & 0.78 & 0.172\footnote{These values are reduced for about 10~\% due to diamagnetic impurities, see text for details.} & 0.44 & 0.41$^a$ & \\
  Pb$_2$VO(PO$_4)_2$ & $-1.9$ & 0.83 & 0.189 & 0.42 & 0.45 & \onlinecite{kaul2004} and \onlinecite{enrique} \\
  SrZnVO(PO$_4)_2$   & $-1.1$ & 0.52 & 0.303 & 0.28 & 0.41 & \onlinecite{enrique} \\
  BaCdVO(PO$_4)_2$   & $-0.9$ & 0.52 & 0.346 & 0.31 & 0.31 & \onlinecite{nath2008} \\
\end{tabular}
\end{ruledtabular}
\end{table*}

Table~\ref{comparison} presents key characteristics of the magnetic susceptibility and the specific heat for a number of FSL materials with FM $\bar J_1$ and AFM $\bar J_2$. These materials reveal different magnitude of the frustration, as evidenced by the effective frustration ratio $\alpha=\bar J_2/\bar J_1$. The $\alpha$ value is reduced from BaCdVO(PO$_4)_2$ to SrZnVO(PO$_4)_2$ and, further, to Pb$_2$VO(PO$_4)_2$ and PbZnVO(PO$_4)_2$. The two latter compounds show similar $\alpha\simeq -1.9$. However, they are different with respect to their NN couplings. The saturation field measurement\cite{high-field} indicates that Pb$_2$VO(PO$_4)_2$ is spatially isotropic  ($J_1\simeq J_1'$), while in the case of PbZnVO(PO$_4)_2$ $J_1'-J_1\simeq 1$~K. Nevertheless, the positions and magnitudes of the susceptibility maxima in the two compounds are close to each other. On the other hand, the enhancement of the frustration in SrZnVO(PO$_4)_2$ and BaCdVO(PO$_4)_2$ leads to the shift of the maximum to lower temperatures and to the increase of the susceptibility value at the maximum. 

At first glance, the characteristics of the specific heat are slightly different: the positions of the maxima show the same trend as observed in the susceptibility data. Yet the maximum specific heat value in PbZnVO(PO$_4)_2$ is lower than that in Pb$_2$VO(PO$_4)_2$ and comparable to $C^{\max}$ in SrZnVO(PO$_4)_2$. Still, this result is likely insignificant, because the $C^{\max}$ value is affected by the impurities. In Sec.~\ref{experiment}, we have shown that both the magnetization and the specific heat data point to $\simeq 10$~\% of diamagnetic impurities in the samples under investigation. Once this error is corrected, we find $C^{\max}\simeq 0.45R$ which coincides with the result for Pb$_2$VO(PO$_4)_2$. Moreover, $\chi^{\max}J_c$ rescales to 0.191 in perfect agreement with $\chi^{\max}J_c=0.189$ in Pb$_2$VO(PO$_4)_2$.

Our comparison of the different FSL materials indicates that the spatial anisotropy of the spin lattice has a minor effect on thermodynamic properties. The behavior of the magnetization and the specific heat is mainly determined by the frustration ratio that controls the magnitude of quantum fluctuations and determines spin correlations in the system. This finding provides experimental verification for simulation results which consistently showed weak changes in thermodynamic properties upon the distortion of the FSL.\cite{tsirlin2009} Thus, the distortion of the spin lattice is a secondary effect compared to the frustration. In the case of the FSL, low-symmetry materials with complex structures can be reliably considered as experimental realizations of the regular model. 

Extending the above statements to the ground state properties and using $\alpha\simeq -1.9$, one would suggest that PbZnVO(PO$_4)_2$ undergoes long-range ordering towards columnar antiferromagnetic state at $T_N\simeq 3.9$~K (in zero field). This suggestion is further confirmed by the reference to Pb$_2$VO(PO$_4)_2$ with similar exchange couplings ($\bar J_1\simeq -5.1$~K, $\bar J_2\simeq 9.4$~K), similar $T_N\simeq 3.5$~K, and similar transition anomalies in the magnetic susceptibility and the specific heat.\cite{kaul2004,enrique} The columnar AFM ordering in Pb$_2$VO(PO$_4)_2$ was directly confirmed by neutron scattering and nuclear magnetic resonance measurements.\cite{skoulatos2009,ramesh2009} Thus, a similar ordering in PbZnVO(PO$_4)_2$ looks likely. Then, the role of the spatial anisotropy is mainly restricted to determine the direction of columns. Since $|J_1'|<|J_1|$, we expect that spins will show parallel alignment along the $b$ direction (interaction $J_1$). To test this prediction, further neutron scattering studies are desirable. We also hope that experimental realizations of the spatially anisotropic FSL will stimulate theoretical investigation of the respective model.

Finally, we turn to the structural aspects of the study. Different geometries of the [VOPO$_4$] layers in the AA$'$VO(PO$_4)_2$ compounds stimulated us to study the geometrical parameters that influence the exchange couplings. For the NNN interactions $J_2$ and $J_2'$, these parameters are the V--O and O--O distances. Mutual orientation of the VO$_5$ pyramids and the PO$_4$ tetrahedron has minor effect on the exchange integrals (within a reasonable range of possible geometries). The role of the tetrahedrally coordinated X cation is to define the size of the tetrahedron and the relevant O--O distance. As the X cation gets larger, the NNN interactions in the FSL materials are reduced. This result helps to explain the change of $J_2$ in the family of the FSL compounds. 

Vanadium phosphates normally show $\bar J_2$ of $9-10$~K\cite{kaul2004,enrique,nath2008} in agreement with the small size of the P$^{+5}$ cation (ionic radius $r=0.17$~\r A,\cite{shannon} typical O--O separations $d_{\text{O--O}}=2.3-2.5$~\r A).\cite{srznvp2o9,baznvp2o9,shpanchenko2006} As we turn to Li$_2$VOSiO$_4$ with the larger Si$^{+4}$ cation ($r=0.26$~\r A,\cite{shannon} $d_{\text{O--O}}=2.63$~\r A\cite{millet1998}), $J_2$ is reduced to $\simeq 6$~K.\cite{rosner2002,enrique} In Li$_2$VOGeO$_4$, one finds an even smaller $J_2\simeq 4$~K\cite{rosner2002,enrique} due to the larger Ge$^{+4}$ cation ($r=0.39$~\r A,\cite{shannon} $d_{\text{O--O}}=2.78$~\r A\cite{millet1998}). We should note that VOMoO$_4$ with \mbox{$J_2\simeq 20$~K}\cite{carretta2002,bombardi2005} does not follow this trend due to the slightly different crystal structure and the different superexchange scenario in this compound. Mo$^{+6}$ is a transition-metal cation, hence its $4d$ orbitals have low energies and show stronger overlap with the vanadium orbitals as compared to the $p$ orbitals of the main-group cations (P, Si, Ge).\cite{carretta2002} Another exception is BaCdVO(PO$_4)_2$ with its unusually low $\bar J_2\simeq 3.2$~K. The origin of this anomaly remains unclear and should be a subject of future investigations. 

Based on the above analysis, we can propose a new route towards strongly frustrated square lattice materials. In the AA$'$VO(PO$_4)_2$, Li$_2$VOSiO$_4$, and Li$_2$VOGeO$_4$ compounds, the $|J_1|<J_2$ regime is realized. To reach the critical regions at $J_2/|J_1|\simeq 0.5$, one has to reduce $J_2$ (irrespective of the sign of $J_1$). Since the $J_2$ value is controlled by the size of the tetrahedron, one should use a larger tetrahedrally coordinated cation. The promising candidate is As$^{+5}$ with $r=0.34$~\r A\cite{shannon} and typical $d_{\text{O--O}}=2.65-2.80$~\r A, providing $J_2$ of about 5~K or even lower. Our preliminary band structure calculations for vanadium arsenates show weak contribution of As states at the Fermi level, hence the magnitude of superexchange interactions will mainly depend on the O--O distances. The chemistry of vanadium(IV) arsenates remains weakly explored (in particular, none of the AA$'$VO(AsO$_4)_2$ compounds have been reported) and demands further investigation.

In summary, we have reported the crystal structure together with an experimental and microscopic study of the magnetic properties of PbZnVO(PO$_4)_2$. This compound can be understood as a frustrated square lattice material with $\bar J_1\simeq -5.2$~K, $\bar J_2\simeq 10.0$~K, and sizable spatial anisotropy of nearest-neighbor couplings ($J_1'/J_1\simeq 0.8$) and next-nearest-neighbor couplings ($J_2'/J_2\simeq 0.7$). The spatial anisotropy shows little effect on thermodynamic properties. At $T_N\simeq 3.9$~K, PbZnVO(PO$_4)_2$ undergoes a phase transition which is likely associated with columnar antiferromagnetic ordering, similar to the predictions of the regular model for the frustration ratio $\alpha=\bar J_2/\bar J_1\simeq -1.9$. Our microscopic study helps to identify the relevant geometrical parameters for the next-nearest-neighbor interactions in FSL materials. We show that the magnitudes of such interactions crucially depend on the size of the tetrahedrally coordinated cation. This finding suggests a new route towards strongly frustrated square lattice materials in the critical region of the respective phase diagram.

\acknowledgments
We are grateful to Klaus Koepernik for implementing Wannier functions in FPLO and to Evgeny Antipov for reading the manuscript. Financial support of RFBR (grant no. 07-03-00890) is acknowledged. A.Ts. also acknowledges the MPI CPfS and the MPI PKS for hospitality and financial support during the stay.


\begin{thebibliography}{53}
\bibitem[{not({\natexlab{a}})}]{note2}
\bibinfo{note}{From the experimental data, we do not find any indications for
  non-stoichiometry of PbZnVO(PO$_4)_2$, despite the reactant mixture was not
  stoichiometric. First, structure refinement leads to reasonable atomic
  displacement parameters (see Table~\ref{coordinates}), hence indicating the
  lack of deficiency in all the positions. Second, magnetization measurements
  do not show any low-temperature upturn and point to the high quality of the
  samples (small amount of magnetic defects). Finally, there is no information
  on the non-stoichiometry of the other AA$'$VO(PO$_4)_2$ compounds. Thus, we
  refer to the regular PbZnVO(PO$_4)_2$ composition.}
\bibitem[{not({\natexlab{b}})}]{note1}
\bibinfo{note}{In low-symmetry structures, the $t_{ij}^{xy\rightarrow\alpha}$
  and $t_{ji}^{\alpha\rightarrow xy}$ hoppings are not equivalent by symmetry.
  Then, both enter the sum with the coefficient 2 instead of 4 in
  Eq.~\eqref{jfm}.}
\bibitem[{not({\natexlab{c}})}]{note3}
\bibinfo{note}{For each polyhedron, we define $\Delta d=d'-d$, where $d$ and
  $d'$ are averaged distances, relevant for $J_2$ and $J_2'$, respectively:
  $d=(d_{\text{T--O(2)}}+d_{\text{T--O(8)}})/2,d'=(d_{\text{T--O(6)}}+d_{\text%
{T--O(9)}})/2$ with T = V and P for the VO$_5$ pyramids and the PO$_4$
  tetrahedra, respectively.}
\end{thebibliography}
\end{document}